\documentclass[aps,prl,twocolumn,superscriptaddress,showpacs,draft]{revtex4}
\usepackage{graphicx}
\input{epsf}

\newcommand{\rem}[1]{}

\begin{document}

\title{Fractal Weyl law for quantum fractal eigenstates}
\author{D.L.Shepelyansky}
\affiliation{\mbox{Laboratoire de Physique Th\'eorique, 
UMR 5152 du CNRS, Universit\'e Toulouse III, 
31062 Toulouse, France}}
%\date{\today}
\date{September  14, 2007}
%05.45.Mt Quantum chaos; semiclassical methods
%05.45.Df Fractals (see also 47.53.+n Fractals in fluid dynamics)
%03.65.Sq Semiclassical theories and applications

%\widetext

\begin{abstract} 
The properties of the resonant Gamow states
are studied numerically in the semiclassical limit for
 the quantum Chirikov standard map with absorption. 
It is shown that the number of such states is described by the fractal 
Weyl law and their Husimi distributions closely follow the strange repeller set
formed by  classical orbits nonescaping  in future times. 
For large matrices the distribution
of escape rates converges to a fixed shape profile 
characterized by a spectral gap related
to the classical escape rate.
\end{abstract}
\maketitle

The Weyl law \cite{weyl} gives a fundamental link between
the properties of quantum eigenstates in closed Hamiltonian systems,
the Planck constant $\hbar$
and the classical phase space volume. 
The number of states in this case is determined by the system dimension
$d$ and the situation is now well understood both on mathematical and 
physical grounds \cite{sjostrand,landau}. Surprisingly, only recently
it has been realized that the case of nonunitary
operators describing open systems in the semiclassical limit
has a number of new interesting properties
and the concept of the fractal Weyl law has been introduced
to describe the dependence of number of resonant Gamow eigenstates 
on $\hbar$ \cite{sjostrand1,zworski2003}.  The Gamow eigenstates
find important applications for decay of radioactive nuclei 
\cite{gamow}, quantum chemistry reactions \cite{moiseyev},
chaotic scattering \cite{gaspard} and microlasers with chaotic resonators
\cite{stone1998,harayama2003,harayama2007}. Thus the understanding
of their properties in the semiclassical limit represents
an important challenge. 

According to the fractal Weyl law \cite{sjostrand1,zworski2003} the number
of Gamow eigenstates $N_\gamma$, which have escape rates $\gamma$
in a finite band width  $0 \leq \gamma \leq \gamma_b$,
scales as
\begin{equation}
N_\gamma \propto \hbar^{-(d-1)}
\label{eq1} 
\end{equation}
where $d$ is a fractal dimension
of a classical strange repeller formed by classical orbits
nonescaping in future (or past) times. By numerical simulations 
it has been shown that the law (\ref{eq1})  works
for a 3-disk system \cite{zworski2003} and quantum chaos maps
with absorption \cite{schomerus,nonnenmacher1} at specific values of $d$.
Recent mathematical results for open quantum maps are presented in
\cite{nonnenmacher2}. The law (\ref{eq1}) for open
systems with a fractal dimension $d<2$ 
leads to a striking consequence: only a relatively
small fraction of eigenstates 
$\mu \sim N_\gamma/N \propto \hbar^{(2-d)}$ 
have finite values of $\gamma$ while almost all
eigenstates of matrix operator of size $N \propto 1/\hbar$
have infinitely large $\gamma$. The later states
are associated \cite{schomerus}
with classical orbits which escapes from the system
after the Ehrenfest time \cite{chirikov1988}. The former states
with finite $\gamma$
are related to the classical fractal repeller
and have been named quantum fractal eigenstates
due to a fractal structure of their Husimi distributions
closely following the classical fractal
\cite{maspero1}. The semiclassical description of
probability density for such states has been 
derived recently in \cite{keating}.
  
In view of the recent results described above 
I study numerically a simple model of 
the quantum Chirikov standard map (kicked rotator)
with absorption introduced in \cite{borgonovi}
which allows to vary continuously the fractal dimension
of the classical strange repeller. In this way
the fractal Weyl law (\ref{eq1}) is verified
in the whole interval $1 \leq d \leq 2$. The model
also allows to establish
the limiting semiclassical distribution  over
escape rates $\gamma$ and find its links with 
the fractal properties of the classical strange repeller.
The Chirikov standard map is a generic model of chaotic dynamics and it 
finds applications in various physical systems
including  magnetic mirror traps, accelerator beams,
Rydberg atoms in a microwave field \cite{chirikov,lichtenberg,ott,hydrogen}.
The quantum model has been built up in experiments 
with cold atoms \cite{raizen}. Thus the results obtained for this
model should be generic and should find  applications for various systems.

The quantum dynamics of the model is described by the evolution matrix:
\begin{eqnarray} 
\bar{\psi} = \hat{U} \psi = \hat{P} e^{-iT\hat{n}^2/4} e^{-ik
\cos{\hat{\theta}}}
 e^{-iT\hat{n}^2/4} \psi,
\label{eq2}
\end{eqnarray}
where $\hat{n}=-i {\partial / { \partial \theta}}$  and 
the operator $\hat{P}$ projects the wave function to the states in the
interval $[ -N/2, N/2 ]$.
The semiclassical limit corresponds to $k \rightarrow \infty$, 
$T \rightarrow 0$ with
the  chaos parameter $K=kT=const$ and 
absorption boundary $a=N/k=const$. Thus $N$ is inversely proportional
to the effective Planck constant $T=\hbar_{eff}$, 
it gives the number of quantum eigenstates
and the number of quantum cells inside the classical phase space.
The classical  dynamics is described by the Chirikov standard map 
\cite{chirikov,lichtenberg} in its symmetric form:
\begin{eqnarray} 
\bar{n} = n + k \sin{ \left[ \theta + { T n \over 2} \right]},
\bar{\theta} = \theta + {T \over 2} (n+\bar{n}).
\label{eq3}
\end{eqnarray}
Physically, the map describes a free particle propagation 
in presence of periodic kicks with period $T$
(e.g. kicks of optical lattice in \cite{raizen}).
In this model all  trajectories 
(and quantum  probabilities) escaping the interval $[ -N/2, N/2 ]$
are absorbed and never return back. It is convenient to fix $K=7$
so that the phase space have no visible stability islands for 
$a \leq 6$ \cite{note1}. Then for the classical dynamics
the probability $P(t)$ to stay inside decays exponentially with time
as $P(t) \sim \exp(-\gamma_c t)$
\cite{borgonovi,maspero2}, where $\gamma_c$ is
the classical escape rate  and $t$ is measured in the number of map iterations.
For large values of $a$ the spreading goes in a diffusive way and
$t_c = 1/\gamma_c \sim  N^2 /D \approx 2 a^2$ where $D \approx k^2 /2$ 
is the diffusion rate for $K \gg 1$. The independence of 
$t_c$ of $N$ implies $a=N/k=const$.
The quantum operator (\ref{eq2}) can be considered as a 
simplified model of chaotic microlasers where 
all rays with orbital momenta below some critical value
determined by the refraction index escape from a microcavity
\cite{stone1998,harayama2003,harayama2007}.
\begin{figure}
\epsfxsize=8.5cm
\epsffile{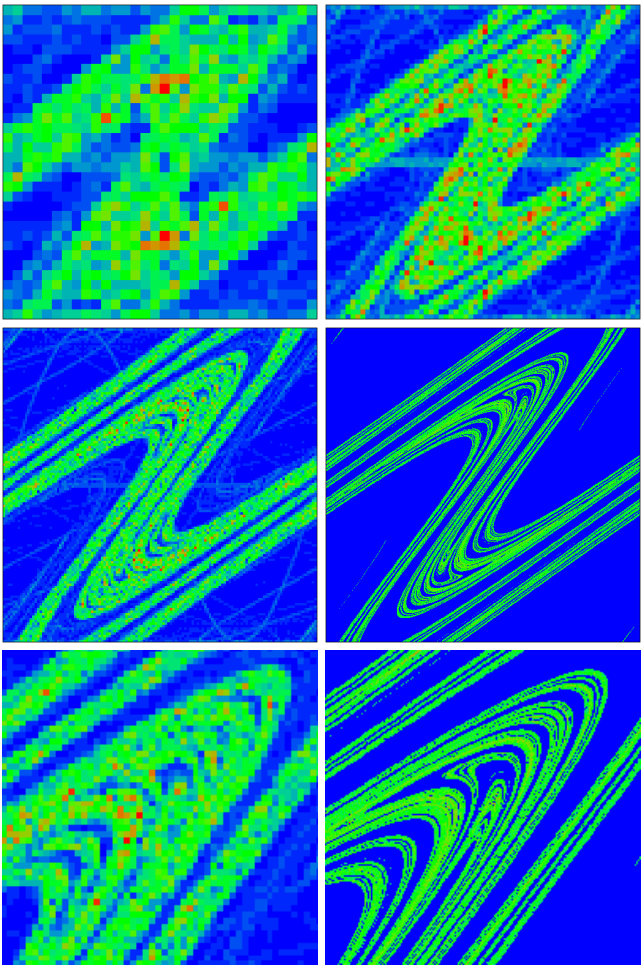}
%\centerline{\epsfxsize=4.2cm\epsffile{fig1a.eps}
%\hfill\epsfxsize=4.2cm\epsffile{fig1b.eps}}
%\vspace{0.1cm}
%\centerline{\epsfxsize=4.2cm\epsffile{fig1c.eps}
%\hfill\epsfxsize=4.2cm\epsffile{fig1d.eps}}
%\vspace{0.1cm}
%\centerline{\epsfxsize=4.2cm\epsffile{fig1e.eps}
%\hfill\epsfxsize=4.2cm\epsffile{fig1f.eps}}
%\vglue -0.0cm
\caption{(color online) Husimi functions of
quantum fractal eigenstates with minimal
value of $\gamma$  at $N=1025$ (top left),
$N=4097$ (top right), $N=16349$ (middle left)
and the density plot of classical strange chaotic repeller 
formed by orbits nonescaping forward in time (middle right); 
two bottom panels show
zoom for two middle panels respectively.
Here $a=N/k=2$, $K=kT=7$ and 
the box counting dimension of the repeller
is $d=1.7230$. In top and middle panels
$0 \leq \theta <2\pi$, $-N/2 \leq n \leq N/2$;
density is proportional to color
with red/gray for maximal density and blue/black for
zero density.
}
\label{fig1}
\end{figure}

The right eigenstates $\psi^{(m)}_n$ and eigenvalues 
$\lambda_m = \exp(-i\epsilon_m - \gamma_m/2)$ 
of the evolution operator $\hat{U}$ 
are determined numerically by direct dioganalization up to
a maximal value $N=22001$ (only states symmetric in $n$ are considered).
The Husimi distribution \cite{husimi}, obtained from smoothing of
a Wigner function on a Planck constant scale,
is shown in Fig.~\ref{fig1} for eigenstates with minimal
$\gamma_m$ at different values of $N$ at $a=2$.
With the increase of $N$ the Husimi distribution
converges to a fractal set which is very similar to the
classical strange repeller formed by classical orbits
never escaping in the future. The classical repeller is obtained
by iterating up to $3 \times 10^{9}$ classical trajectories
homogeneously distributed in the whole phase space
at $t=0$. The classical remaining probability  $P(t)$
decays exponentially with $\gamma_c=0.2702 \pm 0.0011$
and the computation of the box counting dimension \cite{lichtenberg,ott} 
of the strange repeller gives $d=1.7230 \pm 0.0085$.
According to \cite{gaspard,lichtenberg,ott}
the information dimension $d_1$ of the repeller can be expressed
as $d_1=2-\gamma_c/\Lambda$, where $\Lambda$ is the Lyapunov exponent.
For  large $a$ and small $\gamma_c$ it
can be expressed via its value for the Hamiltonian dynamics on a torus
where $\Lambda \approx \ln (K/2) = 1.2527$ (for $K=7$)
\cite{chirikov}. This gives $d_1=1.7843$
that is rather close to the numerical value
of box counting dimension $d$ (usually these two dimensions are rather close 
and, contrary to \cite{nonnenmacher1}, 
I will not make difference between them).
However, for smaller values of $a$ 
the relation  $\Lambda = \ln (K/2)$ is no more valid.
To have $\Lambda$ for all values of $a$ 
its value is computed numerically following 
approximation used in \cite{chirikov}:
$\Lambda = \langle  \mid \ln (K\cos(\theta+Tn/2)) \mid \rangle$
where the average $\langle...\rangle$ is done over the orbits on the repeller.
In this way $\Lambda$ varies in the interval
$1.913 \leq \Lambda \leq 1.294$ for $0.7 \leq a \leq 6$
($\Lambda=1.363$ and $d_1=1.801$ at $a=2$).
\begin{figure}
\centerline{\epsfxsize=7.5cm\epsffile{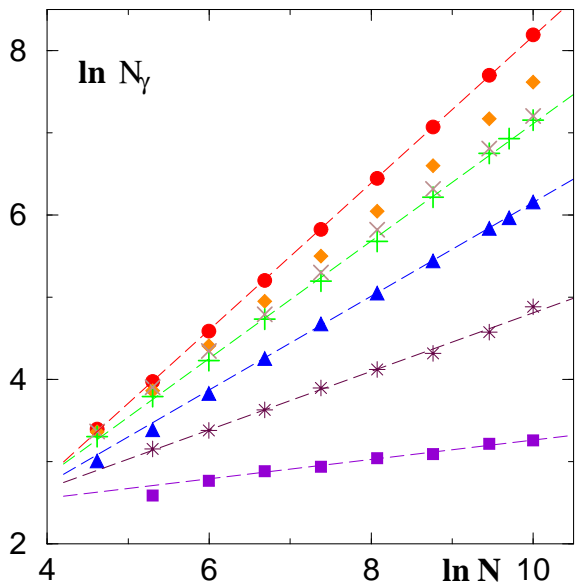}}
\vglue -0.2cm
\caption{(color online) Dependence of 
the integrated number of states $N_\gamma$
with escape rates $\gamma \leq \gamma_{b}=8/a^2$ 
on matrix size $N$.
Symbols show numerical data for various values of absorption border 
$a=N/k$: 4 (circles), 2.5 (diamonds), 2 (+), 1.5 ($\times$),
1 (triangles), 0.8 (*), 0.7 (squares).
Dashed lines show algebraic fits
$N_\gamma \propto N^\nu$ with the fractal Weyl exponent 
$\nu=0.8930 \pm 0.0028$ ($a=4$), 
$\nu=0.7129 \pm 0.0073$ ($a=2$),
$\nu=0.5697 \pm 0.0042$ ($a=1$),
$\nu=0.3559 \pm 0.0094$ ($a=0.8$)
$\nu=0.1175 \pm 0.0069$ ($a=0.7$, here $\gamma_{b}=4/a^2$).
Logarithms are natural.
}
\label{fig2}
\end{figure}

To check the validity of the fractal Weyl law (\ref{eq1})
for various $d$ the absorption border $a$
is varied in the interval $0.7 \leq a \leq 6$
so that the classical fractal dimension and decay rate
vary in the intervals
$0.9976 \pm 0.0060 \leq d \leq 1.9367 \pm 0.0067$
and $1.6349 \pm 0.0135 \leq \gamma_c \leq 0.0592 \pm 0.0003$.
In the quantum case the number of
states $N_\gamma$ is computed
in the band width $0 \leq \gamma \leq \gamma_b$
with $\gamma_b=8/a^2$. 
In this way $\gamma_b > \gamma_c$ and the band contains
a large fraction of fractal eigenstates.
To improve the statistics,
$N_\gamma$ is averaged over $N_r$
cases with slightly different values
of $k \pm \delta k$ with $\delta k \leq 2$.
Such a small variation of $k$ does not affect the semiclassical
properties but allows to improve the statistical accuracy.
The number of realizations varied from $N_r=40$
at $N=101$ to $N_r=1$ at $N=22001$.
The dependence of integrated number of states $N_\gamma$
on $N \propto 1/\hbar$ is shown in Fig.~\ref{fig2}.
The fit $N_\gamma \propto N^\nu$
allows to determine the exponent 
$\nu$ which according to (\ref{eq1})
should satisfy the relation $\nu=d-1$.
It is important to stress that  
 $N_\gamma \ll N/2$ at $a \leq 2$,
so that the main part of $(N+1)/2$ eigenvalues has enormously
large $\gamma \gg \gamma_b$.
\begin{figure}
\centerline{\epsfxsize=7.5cm\epsffile{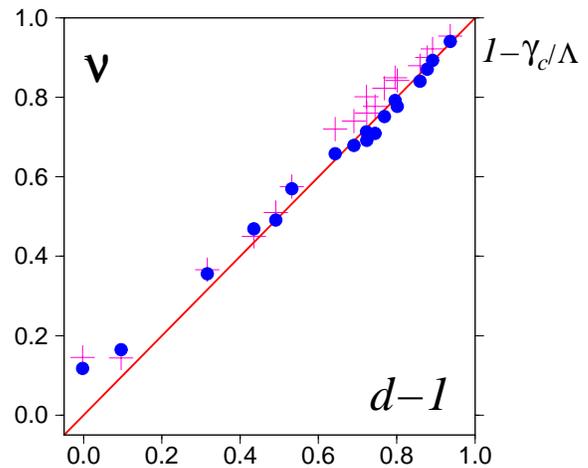}}
\vglue -0.2cm
\caption{(color online) Dependence of 
the fractal Weyl exponent $\nu$ on the fractal box
counting dimension $d$: full circles show numerical data,
the straight line shows the fractal Weyl law (\ref{eq1}) $\nu=d - 1$,
pluses show  $1-\gamma_c/\Lambda$ versus
$d-1$ which should follow the relation 
$d_1 -1 = 1-\gamma_c/\Lambda$,
where $\Lambda$ is the Lyaponov exponent
computed approximately (see text).
Here $0.7 \leq a \leq 6$,
$0.1175 \leq \nu \leq 0.9402$,
$0.9976 \leq d \leq  1.9367$.
}
\label{fig3}
\end{figure}

The dependence of $\nu$ on $d$ is shown in Fig.~\ref{fig3}. 
The law (\ref{eq1}) is well satisfied for fractal dimensions
$1 \leq d < 2$. Certain deviations for $d$ close to $1$
should be attributed to rather small values
of $N_\gamma$ ({\it e.g.} $N_\gamma=26$ at $a=0.7$ and $N=22001$)
so that even larger $N$ values are required to 
see the asymptotic behavior. The relation 
$\nu=1-\gamma_c/\Lambda$ also works 
rather well even if 
the Lyapunov exponent $\Lambda$
should be probably computed in a more exact way
for small values of $a \sim 0.7$.
Thus, the data of Fig.~\ref{fig3} confirms the validity
of the fractal Weyl law in the whole available interval
of fractal dimensions.
\begin{figure}
\centerline{\epsfxsize=7.5cm\epsffile{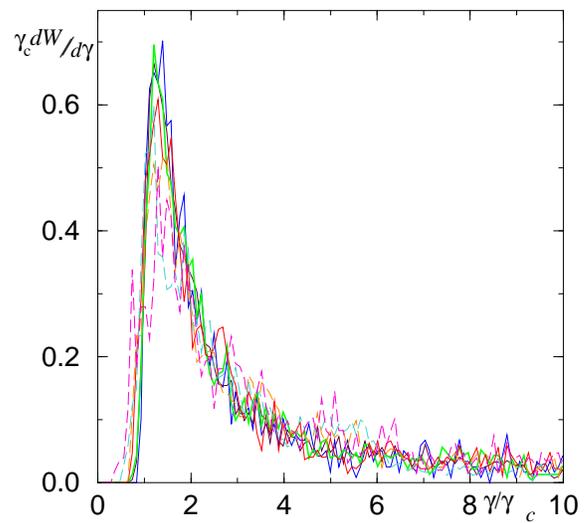}}
\vglue -0.2cm
\caption{(color online) 
Dependence of the distribution
$\gamma_c d W/d \gamma$ on
the rescaled escape rate $\gamma/\gamma_c$
for different values of $N$ at $a=2$.
Here $\gamma_c=0.2702$, $d=1.7230$,
the probability $d W/d \gamma$ is normalized to unity
in the interval $\gamma_b$ and
$N$ is 22001 (blue/black full curve $N_\gamma=1278$),
12801 (maroon/gray full $N_\gamma=1022$), 
6401 (green/gray full $N_\gamma=500$), 
3201 (red/gray full $N_\gamma=293$),
1601 (orange/gray dashed $N_\gamma=181$), 
801 (turqse/gray dashed $N_\gamma=114$),
401 (magenta/gray dashed $N_\gamma=68.7$).
}
\label{fig4}
\end{figure}

In addition to the integrated characteristic (\ref{eq1})
it is interesting to consider the differential
distribution $d W/d\gamma$ which determines
the number of states in the interval $d\gamma$
at given $\gamma$. The evolution of distributions $dW/d\gamma$
with the growth of $N$ is shown in Fig.~\ref{fig4}.
The data clearly show that in the semiclassical limit 
$dW/d\gamma$ converges to a certain 
limiting distribution independent of $N$.
Such an effect has been noticed already in 
first studies \cite{borgonovi} where mainly the diffusive
limit with $a=10$ and $\gamma_c \ll \Lambda$
has been considered. In such a case the dimension is
very close to the integer value $d=2$
and due to that the fractal dependence
(\ref{eq1}) has been missed in
\cite{borgonovi,maspero1}
even if the fractal structure of eigenstates
has been clearly detected \cite{maspero1}.
In the diffusive case $d\approx2$
one has $dW/d\gamma \propto 1/\gamma^{3/2}$
for $\gamma > \gamma_c$
that is explained by simple estimates \cite{borgonovi}
and more rigorous analytical treatment \cite{skipetrov}.
When the fractal dimension $d$ is noticeably less than 2 
than $\gamma_c \sim \Lambda$ and the diffusive 
approximation is no more valid. A distinctive feature of the distribution
in this case is the gap in the distribution $dW/d\gamma$
which is zero for $\gamma < \gamma_c$, sharp peak at $\gamma = \gamma_c$
followed by a smooth drop at $\gamma > \gamma_c$
(this drop is compatible with dependence $1/\gamma^{3/2}$). 
\begin{figure}
\centerline{\epsfxsize=7.5cm\epsffile{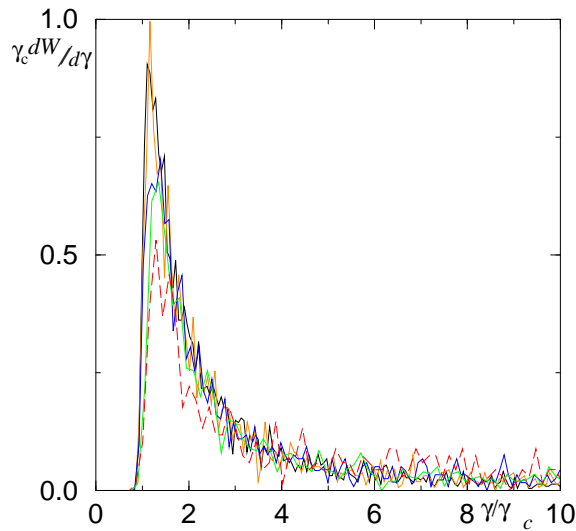}}
\vglue -0.2cm
\caption{(color online) 
Same distribution as in Fig.~\ref{fig4}
drown for various absorption boundaries $a$
at $N=22001$. Here $a$ is
4 ($\gamma_c=0.1019$, $N_\gamma=3607$, black curve);
2.5 ($\gamma_c=0.2063$, $N_\gamma =2032$, orange/gray curve);
2 ($\gamma_c=0.2702$, $N_\gamma=1278$, blue/black curve);
1.5 ($\gamma_c=0.2961$, $N_\gamma=1342$, green/gray curve);
1 ($\gamma_c=0.6967$, $N_\gamma=472$, red/gray dashed curve).
}
\label{fig5}
\end{figure}

These properties of the distribution $dW/d\gamma$
remain essentially the same when $\gamma_c$ is changed by
a factor 3.5 as it is shown in Fig.~\ref{fig5}. Indeed, 
the shape of the distribution
varies very little for $1.5 \leq a \leq 4$
and becomes broader only at $a<1.5$ . The later case have 
however relatively small statistics $N_\gamma$
and probably larger $N$ should be used to
reach a limiting distribution for $a < 1.5$.
It is interesting to note that  $dW/d\gamma$
has certain similarities with 
the Wigner proper times distribution discussed
in \cite{brouwer}.

In conclusion, the obtained data 
confirm the validity of the fractal Weyl
for all fractal dimensions in the interval $1 \leq d \leq 2$.
They show the existence of the limiting distribution
of the Gamow resonances $dW/d\gamma$
which has a gap of size $\gamma_c$ 
above which the distribution has a sharp peak
(see Figs.~\ref{fig4},\ref{fig5}). Thus the classical
decay rate $\gamma_c$ essentially determines the quantum decay
rates on the quantum fractal corresponding to 
the classical strange repeller with orbits never escaping in 
future times (Fig.~\ref{fig1}). The analytical computation of the
limiting distribution  $dW/d\gamma$ still remains an open problem.
It is possible that the analytical methods
pushed forward recently \cite{keating}
will allow to make progress in this direction.
Also, it would be interesting to check the validity of the fractal Weyl
law for dimensions $d>2$. 
In such a case
it is natural to expect that $N_\gamma \propto \hbar^{-\nu}$
with $\nu = d- n_f$ where $d$ is the fractal dimension of the 
classical strange repeller and $n_f$ is the number of 
degrees of freedom (in the present model $n_f=1, d \leq2$).

At present the properties of large nonunitary matrices
find important applications in various areas including
search on the Internet \cite{google,google1}
and it is possible that the fractal quantum eigenstates
may have there certain applications
since they give an example of important
nontrivially connected fractal sets of small measure.

I thank S.~Nonnenmacher for highlighting of his results and K.M.~Frahm
for stimulating discussions.

%%%%%%%%%%%%%%%%%%%%%%%%%%%%%%%%%%%%%%%%%%%%%%%%%%%%%%%%%%%%%%%%%%%

\end{document}